\documentclass{PoS}

\newcommand{\be}{\begin{equation}}
\newcommand{\ee}{\end{equation}}
\newcommand{\bea}{\begin{eqnarray}}
\newcommand{\eea}{\end{eqnarray}}
\newcommand{\ba}{\begin{array}}
\newcommand{\ea}{\end{array}}
\newcommand{\nn}{\nonumber \\}

\def\hf{\hat{f}}

\def\hF{\hat{F}}
\def\hf{\hat{f}}
\def\hA{\hat{A}}

\def\be{\begin{equation}}
\def\ee{\end{equation}}

\def\fr{\frac}
\def\a{\alpha}
\def\b{\beta}

\def\d{\delta}
\def\e{\epsilon}

\def\l{\lambda}
\def\m{\mu}
\def\n{\nu}
\def\n{\nu}
\def\r{\rho}

\def\d{\delta}

\def\L{\Lambda}

\def\p{\partial}

\def\nn{\noindent}
\def\no{\nonumber}

  \def\cL{{\cal L}}

  \def\cU{{\cal U}}

\title{Duality in Noncommutative Maxwell-Chern-Simons Theory}

\ShortTitle{Duality in NC MCS Theory}

\author{\speaker{Victor O. Rivelles} \\
        Instituto de F\'\i sica, Universidade de S\~ao Paulo\\
        Caixa Postal 66318, 05315-970, S\~ao Paulo, SP, Brazil  \\
        E-mail: \email{rivelles@fma.if.usp.br}}

\abstract{  Applying a master action technique we obtain the dual of the   
noncommutative Maxwell-Chern-Simons theory. The 
equivalence between the Maxwell-Chern-Simons theory and the self-dual
model in commutative space-time does not survive in the
non-commutative setting. We also point out an ambiguity in the
Seiberg-Witten map.         }

\FullConference{Fifth International Conference on Mathematical Methods
  in Physics --- IC2006 \\ 
  April 24-28 2006\\ 
  Centro Brasileiro de Pesquisas Fisicas, Rio de Janeiro, Brazil}

\begin{document}

I will report on a work done in collaboration with E. Harikumar about
duality in noncommutative theories in three dimensions
\cite{Harikumar:2005ry}. The
generalization of the well known equivalence between the
Maxwell-Chern-Simons (MCS) theory and the self-dual (SD) model
\cite{mcssd} to NC space-time was investigated in
\cite{Harikumar:2005ry}. The master
action technique, which was used to establish the equivalence between
these models in commutative space-time, has been adopted in \cite{sg}
and \cite{botta} and these authors have reached different conclusions
regarding the equivalence in the NC setting. In \cite{botta}, after
eliminating some of the fields from the master action, the
perturbative solution to the field equations were used and it was
argued that the NCMCS theory is equivalent to the NCSD model when the
Chern-Simons (CS) term has a cubic contribution like in the
non-Abelian case.  In \cite{sg}, however, which also used the master
action method, it was argued that the NCMCS theory constructed by
applying the inverse Seiberg-Witten (SW) map \cite{sw}, is equivalent
to a theory where the cubic interaction of the vector field is absent
in the CS term. A different approach to study the equivalence has been
adopted in \cite{cw}. Using an iterative embedding method \cite{cwjrf}
for the NCSD model, a dual equivalent theory was constructed to all
orders in the NC parameter. This dual model differs from NCMCS theory
in the coefficient of the cubic interaction of the CS term and this
breaks gauge invariance.  In \cite{davi}, the SW mapped NCMCS theory
was argued to be equivalent to a theory where the effect of
noncommutativity appears through a non-covariant term. This term vanishes
in the commutative limit  and the SD model is then recovered. It is
then imperative, using alternative approaches, to reexamine the
relation between NCMCS theory and NCSD model since the previous
studies are inconclusive. Also, this result has interesting
implications for deriving the bosonization rules for the NC massive
Thirring model \cite{sg,botta}.

Here we will use a procedure which was applied to get a dual
description of the sigma model \cite{bp} and was also used recently
to show the equivalence between massive Abelian gauge theories in
$3+1$ dimensions \cite{hm}. We first apply the procedure to the
partition function of the SW mapped NCMCS theory to order $\theta$ and
derive the dual theory also to order $\theta$. We then argue that this
result can be extended to all orders in $\theta$. From the dual theory
constructed, we show that the equivalence between the MCS theory and
the SD model do not get generalized to the NC setting. In our way to
derive the SW map for the NCMCS theory we found that the presence of a
massive coupling constant turns the map ambiguous. An infinite number
of terms can be present in the map but we choose the minimal set
required by the map.

\section{Ambiguity in the Seiberg-Witten Map}

The SW map is obtained by requiring that an ordinary gauge
transformation on $A_\mu$ with parameter $\lambda$ is
equivalent to a NC gauge transformation on $\hat{A}_\mu$ with gauge
parameter 
$\hat{\lambda}$ so that ordinary gauge fields that are gauge equivalent
are mapped into NC gauge fields that are also equivalent. In four
dimension, where it was originally derived, the SW map for the Abelian
gauge theory to first order in $\theta$ is given by 
\bea
{\hat A}_\m&=&A_\m-\fr{1}{2}\theta^{\a\b}A_\a(2\p_\b A_\m-\p_\m
A_\b), \label{SWmapA}\\
{\hat \l}&=&\l+\fr{1}{2}\theta^{\a\b}\p_\a\l A_\b. 
\label{SWmapL}
\eea
The NC action, when expanded to first order in $\theta$, 
\be
\label{NCaction}
\hat{S} = -\frac{1}{4} \int d^4x \,\, \hat{f}^{\mu\nu} (
\hat{f}_{\mu\nu} + 2 \theta^{\alpha\beta} \partial_\alpha \hat{A}_\mu
\partial_\beta \hat{A}_\nu ),
\ee
with $\hat{f}_{\mu\nu} = \partial_\mu \hat{A}_\nu - \partial_\nu
\hat{A}_\mu$, gives rise to the SW action 
\be
\label{SW4D}
S_{SW} = -\frac{1}{4} \int d^4x \,\, \left[ f^2 + 2
  \theta^{\alpha\beta} ( f^{\mu\nu} f_{\mu\alpha} f_{\nu\beta} -
  \frac{1}{4} f_{\alpha\beta} f^2  ) \right].
\ee
The question we are interested in is the freedom allowed by the SW
  map. Due to its nature we can add to the map (\ref{SWmapA}) any gauge
  invariant term built with $\theta$ and derivatives of the gauge
  field with the right dimension and the new map will still
  be a SW map. The question is then how the SW action will be
  affected. To answer this question let us note that by adding to the
  map (\ref{SWmapA}) a term like 
\be
\delta \hat{A}_\mu = \theta^{\alpha\beta} T_{\mu\alpha\beta}, 
\ee
we get a contribution to the action (\ref{SW4D}) like 
\be
\label{integral4}
\delta \hat{S} = - \int d^4x \,\, \theta^{\alpha\beta} f^{\mu\nu}
  \partial_\mu T_{\nu\alpha\beta}.
\ee
Then if this integral vanishes we will not get any new contribution to
  the SW action. Since in four dimensions the gauge
  field has dimension one the only gauge invariant terms we can add to
  the SW map have $T_{\mu\alpha\beta}$ of the form  $\partial_\mu
  f_{\alpha\beta}$, $\partial_\alpha f_{\mu\beta}$ and $\partial^\r
  f_{\r\b}\eta_{\a\m}$.  
The first term is a gauge transformation to order $\theta$ \cite{taik} 
and gives no contribution to the SW action. The second one is
  proportional the first after applying 
the Bianchi identity. Finally, the third term gives no contribution to
the action since the integral in (\ref{integral4}) vanishes. Then the
SW map to order $\theta$ is essentially unique in four
dimensions. However, as we shall see, in three dimensions the
situation is completely different.

In three dimensions the NCMCS theory is described by the Lagrangian
\be
\hat{{\cL}}_{NCMCS}=-\fr{1}{4g^2}\hF_{\m\n} *\hF^{\m\n}
+\fr{\m}{2}\e_{\m\n\l}\hA^\m*(\hF^{\n\l}+\fr{2i}{3}\hA^\n *\hA^\l),
\label{ncmcs}
\ee
where $\hF_{\m\n}=\p\hA_\m-\p\hA_\n -i[\hA_\m,\hA_\n]_{*}$ while the
NCSD model with a compensating St\"uckelberg field has a Lagrangian
given by
\be
\label{NCsd}
\hat{{\cL}}_{NCSD}=\fr{g^2}{2}({\hf}_\m-{\hat b}_\m)*
(\hf^\m- {\hat b}^\m)
-\fr{1}{2k}\e_{\m\n\l}{\hf}^\m *(\p^\n \hf^\l-\fr{2i}{3}\hf^\n *\hf^\l),
\ee
where ${\hat b}_\m=i{\hat {\cU}}^{-1}*\p_\m{\hat
{\cU}},~{\hat{\cU}}\in U(1)$. 
The NCMCS theory is invariant under the $U(1)$ gauge transformation
\be
\label{NCmcs}
\hA_\m\to {\hat U}^{-1}*\hA_\m *{\hat U}+i{\hat U}^{-1}*\p_\m{\hat U},
\ee
while the NC St\"uckelberg-SD Lagrangian is invariant under
\bea
\hf_\m&\to& {\hat U}^{-1}*\hf_\m *{\hat U}+i{\hat U}^{-1}*\p_\m{\hat U},\no\\
{\hat{\cU}}&\to&{\hat{\cU}}*{\hat U}.
\label{ncu1}
\eea

We should remark that for the pure NCCS theory the SW map has the form
(\ref{SWmapA}) if the CS 
coefficient $\mu$ is chosen to be dimensionless so that the gauge field has
dimension one. The pure NCCS theory has the remarkable property that the
SW action has no dependence whatsoever in $\theta$  \cite{gran}.

In the NCMCS theory and NCSD model the situation is rather different
since one of the
couplings must be dimensionfull and this choice determines the gauge
field dimensionality. If we make the usual choice for the gauge field
dimensionality to be one then $g^2$ in the NCMCS theory has dimension
one and we can use the SW map (\ref{SWmapA}) to obtain 
\be
\label{SWaction}
{\cL}_{SW}=-\fr{1}{4g^2}\left[
F_{\m\n}F^{\m\n}+2\theta^{\a\b}F_{\a\m}F_{\b\n}F^{\m\n}-\fr{1}{2}\theta^{\a\b}
F_{\a\b}F_{\m\n}F^{\m\n}\right] + \frac{\mu}{4} \e^{\m\n\l}A_\m
F_{\nu\lambda}. 
\ee
The fact that $g^2$ has dimension one 
means now that the SW map (\ref{SWmapA}) has an arbitrariness since we can
add an infinite number of gauge invariant terms, all linear in 
$\theta$, but with different powers of derivatives of
$F_{\mu\nu}$. These arbitrary terms in the SW map have the form $g^6 
\theta^{\alpha\beta} T_{\mu\alpha\beta}$ where the $g^6$ factor was
chosen so that $T_{\mu\alpha\beta}$
is a dimensionless function of  $F_{\mu\nu}$ and its derivatives times an
appropriate power of $g$.  We should then ask  whether such terms
contribute to the SW action (\ref{SWaction}). We find that their
contribution has the form 
\be
\label{lambda}
\int d^3x \,\, F^{\mu\nu} ( \partial_{\mu} T_{\nu\alpha\beta} -  \mu g^2
  \epsilon_{\mu\nu\rho} {T^\rho}_{\alpha\beta} ).
\ee

Let us now examine the first terms in the expansion of
$T_{\mu\alpha\beta}$ in powers of $1/g$. The leading terms are
\be
\frac{1}{g^4} \epsilon_{\alpha\beta\rho}{F^\rho}_\mu, \qquad 
\frac{1}{g^4} {\epsilon_{\mu[\alpha}}^\rho F_{\beta]\rho}.
\ee
The first term can be removed by a gauge transformation and a rigid
translation while for the second term (\ref{lambda}) vanishes so both
can be disregarded. The next terms have the form
\be
\frac{1}{g^6} \partial_\mu F_{\alpha\beta}, \qquad 
\frac{1}{g^6} \partial_{[\alpha} F_{\beta]\mu},
\ee
and again the first term can be removed by a gauge transformation
while the second is proportional to the first after using the Bianchi
identity. Higher order terms, however, can contribute. For
instance, to order $1/g^8$ we find that $\epsilon_{\mu\alpha\beta}
F^2$ gives a non trivial contribution since (\ref{lambda}) does not
vanish. Its contribution to the SW
action (\ref{SWaction}) is  
\be
-\frac{1}{g^4} \theta^{\alpha\beta} \epsilon_{\alpha\beta\mu} F^2 
\partial_\nu F^{\mu\nu} - \frac{2\mu}{g^2} \theta^{\alpha\beta}
F_{\alpha\beta} F^{\mu\nu} F_{\mu\nu}.
\ee
Notice that we get a contribution of order $1/g^2$ and the coefficient
of such a contribution could be
chosen to cancel the corresponding term in (\ref{SWaction}).  

The ambiguity found here is not of the same sort as that found by
successive applications of the SW map \cite{taik}. Here it arises
because the model has a dimensionfull coupling constant. If we
require the SW map to be universal in the sense that it applies to any
gauge theory then such terms are not present. We will take this
point of view from now on.  

In \cite{nel} the SW map for the NC St\"uckelberg-Proca
theory has been obtained by requiring that in the unitary
gauge it gives the Proca theory. Using the same criterion, the SW map
for the NC St\"uckelberg-SD model is found to be 
\bea
{\hat f}_\m&=&f_\m-\fr{1}{2}\theta^{\a\b}b_\a(2\p_\b f_\m-\p_\m b_\b),\no\\
{\hat b}_\m&=&b_\m+\fr{1}{2}\theta^{\a\b}\p_\a b_\m b_\b, 
\label{swmssd}
\eea
while the gauge parameter transforms as
\be
{\hat \alpha} = \alpha - \frac{1}{2} \theta^{\alpha\beta} b_\alpha
\partial_\beta \alpha.
\ee
Applying the map to (\ref{NCsd}) we obtain the SW mapped action
\bea
L_{SWSD}&=&\int d^3 x
\fr{g^2}{2}\left[(f_\m-b_\m)(f^\m-b^\m)+\theta^{\a\b}(f_\m-b_\m)
(2b_\a\p_\b f_\m-b_\a\p_\m b_\b+\p_\a b_\m b_\b)\right]\no\\
&-&\fr{1}{4k}\int d^3x \e_{\m\n\l}f^{\m\n}f^\l
-\theta^{\a\b}\e_{\m\n\l}\left[f^{\m\n}b_\a(2\p_\b f^\l-\p^\m
b_\b)+\fr{4}{3}f^\m\p_\a f^\n \p_\b f^\l\right].
\label{swmsd}
\eea

\section{Equivalence of the MCS theory and the SD model}

In order to make the procedure of deriving the dual theory 
in NC space-time more transparent and also to set up our notation, we
present a brief derivation of the well known equivalence between the
MCS theory and the SD model in commutative space-time. The MCS theory
described by the Lagrangian 
\be
{\cL}_{MCS}=-\fr{1}{4g^2}F_{\m\n}F^{\m\n}+\fr{\m}{2}\e_{\m\n\l}A^\m\p^\n
A^\l, 
\label{mcs1}
\ee
is invariant under the $U(1)$ gauge transformation $A_\m\to
A_\m+\p_\m\a$ while the SD model, whose  Lagrangian is 
\be
{\cL}_{SD}=\fr{g^2}{2}f_\m f^\m -\fr{1}{2k}\e_{\m\n\l}f^\m \p^\n f^\l, 
\label{sd}
\ee
has no such an invariance since the $f_\m f^\m$ term breaks the
symmetry. Their equivalence has been analyzed using a phase space path
integral approach \cite{rab} and it was shown that the SD model is
equivalent to a gauge fixed version of MCS theory. Also, this
equivalence has been been studied within the generalized canonical
framework of Batalin and Fradkin in \cite{rab1}. It was shown that the
gauge invariant formulation obtained by the Hamiltonian embedding of
SD model is equivalent to the $U(1)$ invariant MCS theory, clarifying
the equivalence between both theories in spite of fact that they have
different gauge structures. The procedure employed here also sheds 
light into this issue as we shall see.

The MCS theory is also invariant under a global shift of the vector
field $A_\m\to A_\m+\xi_\m$ apart from the $U(1)$ gauge invariance. 
We first elevate this global shift symmetry to a
local one by gauging it by an appropriate antisymmetric gauge field
$G_{\mu\nu}$ which transforms as $G_{\mu\nu} \to G_{\mu\nu} +
\partial_\mu \xi_\nu - \partial_\nu \xi_\mu$. To have the same
physical content as our starting MCS theory we then constrain this
gauge field to be non-propagating. This is done by introducing a
Lagrange multiplier $\Phi$ which imposes the dual field strength of
this gauge field to be flat. The result is
\bea
{\cL}&=&-\fr{1}{4g^2}(F_{\m\n}-G_{\m\n})(F^{\m\n}-G^{\m\n})+\fr{\m}{4}\e_{\m\n\l}P^\m (F^{\n\l}-G^{\n\l})
-\fr{\m}{8}\e_{\m\n\l}P^\m\p^\n P^\l\no\\
&+&\fr{1}{4}\e_{\m\n\l}G^{\m\n}\p^\l\Phi
+\fr{1}{4}\e_{\m\n\l}J^\m(F^{\n\l}-G^{\n\l}),
\label{msc}
\eea
where we have introduced an auxiliary field $P_\m$ to linearize
the CS term. This field has a $U(1)$ gauge invariance 
$P_\m\to P_\m+\p_\m \chi$ when the multiplier field transforms as
$\Phi\to \Phi+\mu\chi$ and $A_\mu \to A_\mu$.  The last term in the
Lagrangian is a source $J^\mu$ 
coupling to the local shift invariant combination of $A_\mu$ and
$G_{\mu\nu}$. The MCS theory 
is recovered from the above Lagrangian by eliminating the $\Phi$ field
using its equation of motion. 

To show the equivalence to the SD model we start from the partition
function 
\be
Z=\int D\Phi DP_\m DA_\m DG_{\m\n} e^{-i\int d^3 x {\cL}}.
\label{part}
\ee
Integrations over $G_{\m\n}$ and $A_\m$ are Gaussian and can be done
trivially leading to 
\be
Z_{dual}=\int D\Phi DP_\m e^{-i\int d^3 x {\cL}_{eff}}.
\label{partd}
\ee
After the redefinitions $\m P_\m = f_\m$ and
$\Phi=\L$, we get the effective Lagrangian 
\be
{\cL}_{eff}=\fr{g^2}{8}(f_\m-\p_\m\L)(f^\m-\p^\m\L)
-\fr{1}{8\mu}\e_{\m\n\l}f^\m\p^\n f^\l +\fr{g^2}{8}J_\m J^\m + 
\fr{g^2}{4}(f^\m-\p^\m\L)J_\m.
\label{dl}
\ee
This theory is invariant under the $U(1)$ gauge transformation
$f_\m\to f_\m+\p_\m\a$ when the St\"uckelberg field transforms as
$\L\to\L+\a$.  We also note that the MCS coupling constant $g^2$ and
the Chern-Simons parameter $\mu$ have both appeared as inverse
couplings when compared with (\ref{sd}).  We can now fix the gauge
invariance in 
(\ref{dl}), for instance by choosing the unitary gauge $\L=0$, to 
recover the self-dual model given in (\ref{sd}). We thus conclude that
the $U(1)$ invariant MCS theory is dual to the $U(1)$ invariant
St\"uckelberg formulation of self-dual model.

From the partition functions (\ref{part}) and (\ref{partd}) 
we derive the mapping between the n-point correlators for 
these theories. For the 2-point function, we get
\be
\left<\e_{\m\n\l}F^{\n\l}(x)~~\e_{\a\b\r}F^{\b\r}(y)\right>\equiv
g^4\left<(f_\m-\p_\m\L)(x)~~(f_\a-\p_\a\L)(y)\right>
+ g^2 g_{\m\a}\d(x-y),
\label{2ptmap}
\ee
leading the identification (up to non-propagating contact terms)
between the gauge invariant combinations 
\be
\e_{\m\n\l}F^{\n\l}\leftrightarrow  g^2(f_\m-\p_\m\L).
\ee

This equivalence between SD model and MCS theory has been extended to
include interaction with matter \cite{cwjrf}. It has been shown that
the SD model minimally coupled to charged dynamical fermionic and
bosonic matter fields is equivalent to a MCS theory non-minimally
coupled to matter. In the weak coupling limit, it was shown in
\cite{nb} that the non-Abelian MCS theory is equivalent to non-Abelian
SD model and recently it was shown that, perturbatively, this
equivalence exists in all regimes of the coupling constant \cite{botta1}.

After re-expressing the NCMCS theory (\ref{ncmcs}) in terms of $A_\m$ and
$\theta^{\a\b}$ using the SW map (\ref{SWmapA}) we apply the above procedure to
construct the corresponding dual theory. Then by comparing this dual
theory with SW mapped NC St\"uckelberg-SD model, we study the status of
their equivalence. We take up this in the next section.

\section{Seiberg-Witten mapped Maxwell-Chern-Simons theory and duality} 

By applying the SW map (\ref{SWmapA}) to the NCMCS
Lagrangian (\ref{ncmcs}) we get to order $\theta$ 
\bea
{\cL}_{SW}&=&-\fr{1}{4g^2}\left[
F_{\m\n}F^{\m\n}+2\theta^{\a\b}F_{\a\m}F_{\b\n}F^{\m\n}-\fr{1}{2}\theta^{\a\b}F_{\a\b}F_{\m\n}F^{\m\n}\right]\no\\
&+&\fr{\m}{4}\e_{\m\n\l}P^\m F^{\n\l}-\fr{\m}{8}\e_{\m\n\l}P^\m\p^\n P^\l,
\label{swmcs}
\eea
where an auxiliary field $P_\mu$ was introduced to linearize the CS
term. We have also used the fact that the
NCCS term gets mapped to the usual commutative CS 
term by the SW map \cite{gran}. After rewriting the above
Lagrangian using auxiliary fields $B_{\m\n}$ and $C_{\m\n}$ as
\bea
{\cL}_{SW}&=&-\fr{1}{4g^2}C_{\m\n}B^{\m\n}
-\fr{\m}{8}\e_{\m\n\l}P^\m\p^\n P^\l+\fr{\m}{4}\e_{\m\n\l}P^\m F^{\n\l}\no\\
&-&\fr{1}{4g^2}\left[
F_{\m\n}F^{\m\n}+2\theta^{\a\b}C_{\a\m}C_{\b\n}F^{\m\n}-\fr{1}{2}
\theta^{\a\b}C_{\a\b}C_{\m\n}F^{\m\n}-B_{\m\n}F^{\m\n}\right],
\label{swmcs1}
\eea
we can now gauge the shift invariance of $A_\m$ field as in the
commutative case. Due to the introduction of $B_{\m\n}$ and $C_{\m\n}$ 
we see that $G_{\m\n}$ will appear quadratically 
and this will simplify the calculation considerably.
So, we introduce a gauge field $G_{\m\n}$ to promote the global
shift invariance of $A_\m$ to a local one. We then get
\bea
{\cL}_{SW}&=&-\fr{1}{4g^2}C_{\m\n}B^{\m\n}-\fr{\m}{2\cdot
4}\e_{\m\n\l}P^\m\p^\n P^\l
+\fr{\m}{4}\e_{\m\n\l}P^\m (F^{\n\l}-G^{\n\l})
-\fr{1}{4}\e_{\m\n\l}G^{\m\n}\p^\l\Phi\no\\
&-&\fr{1}{4g^2}\left[
(F_{\m\n}-G_{\m\n})+(2\theta^{\a\b}C_{\a\m}C_{\b\n}-\fr{1}{2}
\theta^{\a\b}C_{\a\b}C_{\m\n})-B_{\m\n}\right](F^{\m\n}-G^{\m\n}).
\label{swmcs11}
\eea

Starting with the partition function 
\be
Z=\int DP_\m D\Phi DC_{\m\n}DB_{\m\n}DA_\m DG_{\m\n} e^{-i\int d^x
  {\cL}_{SW}}, 
\ee
we can integrate over $G_{\m\n}$, $A_\m$  and $B_{\m\n}$ 
to get the partition function corresponding to the
effective Lagrangian
\bea
{\cL}_{eff}&=&-\fr{\m}{8}\e_{\m\n\l}P^\m\p^\n P^\l
-\fr{1}{4g^2}C_{\m\n}C^{\m\n}+\fr{1}{4}\e_{\m\n\l}C^{\m\n}(\mu P^\l - \p^\l
\Phi) \no \\  
&-&\fr{1}{4g^2}C^{\m\n}\left[ 2\theta^{\a\b}C_{\a\m}C_{\b\n}-\fr{1}{2}
\theta^{\a\b}C_{\a\b}C_{\m\n}\right].
\eea
We have neglected higher order terms in $\theta$ in performing the
Gaussian integrals. It is easy to see that in the commutative limit we
get (\ref{sd}) when $C_{\m\n}$ is eliminated by using its field
equation and setting $\Phi=0$. 

In the NC case $C_{\m\n}$ can be eliminated perturbatively in $\theta$. 
We then get 
\bea
{\cL}_{dual}&=&\fr{g^2}{8}(f_\m-\p_\m\L)(f^\m-\p^\m\L)
+\fr{g^4}{32}\theta^{\a\b}\e_{\a\b\l}(f^\l-\p^\l\L)(f^\m-\p^\m\L)
(f_\m-\p_\m\L)\no\\
&-&\fr{1}{8\mu}\e_{\m\n\l}f^\m\p^\n f^\l,
\label{dswmcs}
\eea
where we have identified $\mu P_\m = f_\m$ and $\Phi =\L$. As in the
commutative case the strong coupling limit of the 
original theory gets mapped into the weak coupling limit of the dual.
It is easy to see 
that in the limit of vanishing $\theta$ the above Lagrangian (in the unitary
gauge where $\L=0$) correctly reproduces the SD Lagrangian (\ref{sd}).

It is interesting to note that the explicit form of the order $\theta$
term in the $C_{\m\n}$ field equation is not need at all to find the
above Lagrangian. This happens
because there are nice cancellations and it is easy to be convinced
that to obtain the dual Lagrangian to $n$-th order in $\theta$ we need the
perturbative solution for $C_{\m\n}$ only to order $(n-1)$.

We can couple a source term $\e_{\m\n\l}F^{\m\n}J^\l$ to the
Lagrangian (\ref{swmcs}) and this leads to the map between 
the 2-point functions  
\bea
&\left<\e_{\m\n\l}F^{\n\l}(x)~~\e_{\a\b\r}F^{\b\r}(y)\right>\equiv
g^4\left<{\tilde f}_\m(x)~~{\tilde f}(y)\right>
+ g^2 g_{\m\a}\d(x-y)&\no\\
&+\fr{g^8}{64}\left< {\bar\theta}_\m{\tilde f}^\n{\tilde f}_\n
+2 {\bar\theta}_\n{\tilde f}^\n{\tilde f}_\m~~
{\bar\theta}_\a{\tilde f}^\b{\tilde f}_\b
+2 {\bar\theta}_\b{\tilde f}^\b{\tilde f}_\a\right>
+g^4( {\bar\theta}_\m{\tilde f}_\a
+{\bar\theta}_\a{\tilde f}_\m +{\bar\theta}_\b{\tilde f}_\b g_{\m\a}),&
\label{2ptncf}
\eea
where ${\bar \theta}_\m=\e_{\m\n\l}\theta^{\n\l}$ and 
${\tilde f}_\m= f_\m-\p_\m\L$. In the limit $\theta\to 0$ we recover
the map obtained in (\ref{2ptmap}).

Here we note that all the $\theta$ dependence of the SW mapped NCMCS
theory comes from the Maxwell term alone as the NCCS term gets 
mapped to usual commutative CS term. Since it is possible to
express the SW mapped Maxwell action to all orders in $\theta$ in terms
of the commutative field strength $F_{\m\n}$ and $\theta$ alone
\cite{gop}(an exact closed form for the SW
mapped Maxwell action is given in \cite{hsy}),
it is easy to convince from (\ref{swmcs}) and (\ref{swmcs11})
that the procedure adopted here can be used to construct the dual theory
to all orders in $\theta$ using a perturbative solution for
the $C_{\m\n}$ field equation.
  
One important point to note is that the theory described by the
Lagrangian (\ref{dswmcs}), which is equivalent to the SW
mapped NCMCS theory, is {\it not} the same as the SW mapped action for
NCSD model (\ref{swmsd}). This clearly shows that the SW mapped
theories are not equivalent.  

\section{Conclusion}\label{con}

In this paper we have constructed and studied the dual description of
the NCMCS theory and investigated the status of the
equivalence between this theory and SD model. 
We have derived the dual theory starting from the SW mapped
NCMCS Lagrangian which is given in terms of commutative fields and the
NC parameter.  The equivalence was obtained at the level of partition
functions and it allowed us to get the mapping between the n-point
correlators of both theories. We have shown that the dual theory does
not coincide with the SW mapped NC St\"uckelberg-SD theory. However, in
the commutative limit, we recover the well known equivalence
between them. We have also shown that the the two-point
correlators map reduces to the one obtained in the commutative case in
this limit. We have argued that this result can be extended to all
orders in $\theta$ due to the structure of the SW mapped NCMCS
Lagrangian. We have also verified that even after accounting for the
ambiguous terms in the SW map, the dual theory and SW mapped NC
St\"uckelberg-SD model are not equivalent.

Hence, we have shown that the equivalence
between the MCS theory and the SD model in commutative space-time does
not survive in the NC case. In this respect we are in agreement with
the results obtained earlier in \cite{sg} and \cite{cw} where it was
argued that these NC theories are not equivalent. But unlike the NCCS
term used in \cite{sg}, we have used the standard NC $U(1)$ invariant
CS term with a cubic interaction as in \cite{cw} and \cite{botta}. The
non-equivalence between the NCSD model and NCMCS theory shown here
will come as an obstacle in generalizing the bosonization of the 
commutative Thirring model to NC space-time as was pointed out in
\cite{sg,cw}.

\vspace{1cm}

\nn{\bf ACKNOWLEDGMENTS}:\\
The work of VOR
was partially supported by CNPq, FAPESP  and PRONEX under contract
CNPq 66.2002/1998-99. 
This paper was done in collaboration with E. Harikumar

\end{document}